\documentstyle[aps,epsf]{revtex}  

\begin{document}        

\baselineskip 14pt
\title{Anisotropies in the CMB}
\author{Martin White}
\address{University of Illinois, Urbana-Champaign}
%
\maketitle              

\begin{abstract}        
        The ten's of micro-Kelvin variations in the temperature of the
        cosmic microwave background (CMB) radiation across the sky encode
        a wealth of information about the Universe.
        The full-sky, high-resolution maps of the CMB that will be made in
        the next decade should determine cosmological parameters to
        unprecedented precision and sharply test inflation and other theories
        of the early Universe.
\
\end{abstract}   	

\section{Introduction}               

Over the last few years it has become common place to speak of the Cosmic
Microwave Background (CMB) anisotropy as the premier laboratory for
cosmology and early universe physics.  In the titles to most talks the
word ``anisotropy'' is often absent, it being understood that the talk
will be about the anisotropy.
This is an interesting phenomenon.  Consider that when we speak of the CMB
we could speak about 3 major properties.
Firstly, the {\it existence\/} of the CMB is one of the pillars of the hot
big bang model of cosmology.  Secondly, the {\it black body\/} spectrum of
the CMB, the most perfect black body ever measured in nature, confirms the
cosmological origin of the CMB and puts extraordinarily strong constraints
on early energy injection in the universe
(e.g.~through decaying particles, see \cite{SmoSco}).
These first two properties show that the CMB has already delivered
important cosmological information.
Our current focus is the third area: the anisotropy.
This fact alone indicates the high level of promise that a study of the
anisotropy holds.

Like much of cosmology, the CMB is a data driven subject.  However, in this
proceedings I focus on the theory behind the CMB anisotropies\footnote{Because
of space, I have referenced primarily work that I have been involved in.
Much more representative referencing can be found in those sources.},
and the current status of theoretical efforts, rather than on the CMB data.
It is a blessing of this field that numerous experimental efforts underway
will make any statements about the experimental situation obsolete before
they reach print (even on the web!).
As an overly brief summary of the current status: the current data are in
good agreement with our general paradigm and support a spatially flat
universe with an almost scale-invariant spectrum of adiabatic fluctuations
in predominantly cold dark matter.
Departures from that statement in any direction fit the data less well,
though at present large error bars and theorist's ingenuity limit the
strength of statements that can be made.

The calculation of CMB anisotropies is now a highly refined subject.  While
most calculations focus on the ``standard'' models, the theory is in fact
very general.  This generality also leads to complexity, but the basic
physics behind the CMB is very simple.
To understand CMB anisotropies it is helpful to recall several general points:
\begin{itemize}
\item The universe was once hot and dense.  At these early times
($\sim 10^5$yr after the bang) the plasma was highly ionized.
Thomson ($\gamma-e$) scattering was rapid and tightly coupled the CMB photons
to the ``baryons'' ($p+e$).  In the limit that this scattering was rapid,
the mean free path was small, a fluid approximation is valid.  Thus we speak
of the photon-baryon fluid.  In this fluid the baryons provide much of the
inertia (mass) and the photons the pressure
($p_\gamma=\rho_\gamma/3$, $p_B\simeq 0$).
\item The observed large-scale structure grew through gravitational
instability from small perturbations at early times.  These density
perturbations imply, through Poisson's equation, small perturbations
in the gravitational field.
\item Combining the above two observations, we infer that the fundamental
modes of the system would be gravity sourced sound waves in the fluid.
The equation of motion for the sound waves can be derived by taking the
tight-coupling limit of the equations of radiative transfer.
In this case (see below)
\begin{equation}
\left[ m_{\rm eff}\ \Delta T_k' \right]' + {k^2 \over 3}\Delta T_k
  = -F_k
\label{eqn:Oscillator}
\end{equation}
where $F_k$ is the gravitational forcing term, $m_{\rm eff}$ describes the
inertia of the fluid, and primes denote derivatives with respect to
(conformal) time.  The forcing term contains derivatives of the potential
(and spatial curvature) while $m_{\rm eff}$ depends on the baryon-to-photon
ratio, which evolves with time.
\item Finally, recombination (when protons captured electrons to form hydrogen
and the universe became neutral) occurred suddenly, but not instantaneously.
With the decrease in the free $e^{-}$ density the mean free path for photons
rises from essentially zero to the size of the observable universe.
The CMB photons travel freely to us, giving us a snapshot of the fluid at a
fixed instant in time.
The energy density, or temperature, fluctuations in the fluid are seen as
CMB temperature differences (anisotropy) across the sky.
\item The temperature fluctuations arise from 3 terms: the gravitational
redshifts as photons climb out of potential wells \cite{SacWol,WhiHu},
density perturbations (with $\Delta T/T=(1/4)\delta\rho_\gamma/\rho_\gamma$)
and Doppler shifts from line-of-sight velocity perturbations.
On large angular scales the first two terms dominate, while on smaller angular
scales the last two are most important.
The density and velocity contributions are out of phase,
with the velocity being smaller than the density contribution (see later).
\end{itemize}

While this way of looking at the anisotropy is physically clear, it is not
how the calculations are actually done.  Remember that the fluctuations are
observed to be small ($\delta\sim 10^{-5}$ c.f.~$\alpha_{QED}\sim 10^{-2}$).
Thus one writes down the Einstein, fluid and radiative transfer equations,
expands about an exact solution and truncates the expansion at linear
order\footnote{The second order terms have been computed and shown to be
small as expected.}.
This procedure gives a set of coupled ODEs which describe the evolution of each
(independent) Fourier\footnote{In hyperbolic geometries the Fourier
decomposition needs to be generalized, but this is a technical point.} mode.
While in some cases the equations can be solved analytically, usually a
numerical solution is performed.

Since the Fourier modes decouple in linear theory it is advantageous to
work in the Fourier basis in the observations also.  Unfortunately the sky
is curved, so plane waves are not the natural basis.  But a ``curved sky
Fourier expansion'' can still be performed using the spherical harmonics:
$\Delta T/T = \sum_\ell a_{\ell m} Y_{\ell m}$.
We focus then not on $\Delta T/T$ but on the $a_{\ell m}$, known as
multipole moments.  By definition $\langle a_{\ell m}\rangle=0$, so the first
non-vanishing correlator is the two-point function.  Since the $Y_{\ell m}$
are a complete orthonormal basis,
$\langle a_{\ell' m'}a_{\ell m}^{*}\rangle\propto\delta_{\ell'\ell}\delta_{m'm}$
and by rotational symmetry the proportionality constants can only depend on
$\ell$.  Thus we write $\langle |a_{\ell m}|^2\rangle=C_\ell$.
If the fluctuations are Gaussian, having specified the mean and variance we
have completely specified the model.  For more general distributions the
higher moments also need to be specified.

\begin{figure}
\begin{center}
\leavevmode
\epsfxsize=8cm \epsfbox{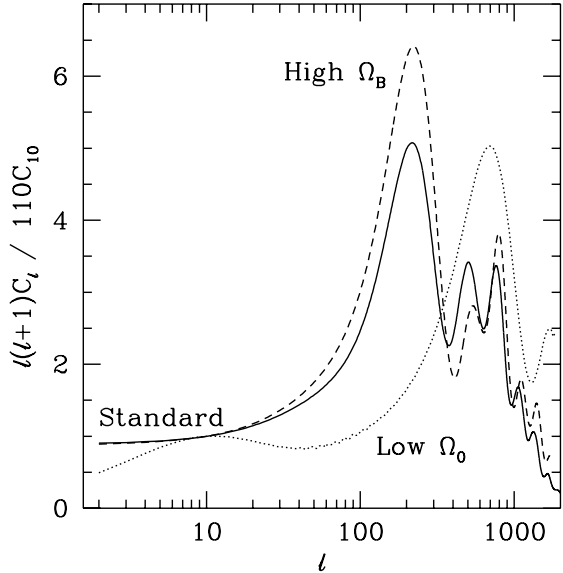}
\end{center}
\caption{The dependence of the angular power spectrum upon some of the
cosmological parameters.  The solid line shows a ``standard'' CDM model.
The dashed line shows the effect of doubling the baryon density (enhancing
the odd peaks, which are waves at maximum compression when the universe
recombined) while the dotted line shows the effect of introducing spatial
curvature (moving features of fixed physical size to smaller angular size,
or higher $\ell$).}
\label{fig:encode}
\end{figure}

We show in Fig.~\ref{fig:encode} a typical $C_\ell$ curve for a standard cold
dark matter (CDM) model.  A readable introduction to the physics can be
found in Refs.~\cite{Romans,PhysTod,Nature,Echoes,ARAA} among others.
The precise shape of the power spectrum depends upon cosmological
parameters as well as the underlying density perturbations and thereby
encodes a wealth of information; see Fig.~\ref{fig:encode}.

The current theoretical situation can be summarized as follows:
\begin{itemize}
\item The formalism for computing $C_\ell$ (and the higher moments)
for any FRW space-time and any model of structure formation exists \cite{HSWZ}.
Since this is essentially a statement in General Relativity, the proof can
be made quite rigorous.
\item Not every model has been calculated, but for those where independent
calculations have been done (mostly CDM models) independent codes agree
to ${\cal O}(1\%)$.
\item The spectrum encodes information on the cosmology and the model of
structure formation and can be measured with exquisite precision.
\item Model dependent parameter extraction can simultaneously fit a dozen
parameters to an accuracy of ${\cal O}(10\%)$ or better, e.g.~\cite{EHT}.
\item This stunning promise has overshadowed an important additional fact
however.  Even if our models do not fit every nuance of the observed data,
model independent constraints on the parameters exist \cite{Signatures} as
do cosmology independent tests of the model of structure formation
\cite{Signatures,TestInf,HSW}.
\end{itemize}

For some time the promise of the CMB to strongly constrain numerous
cosmological parameters has been evident.  Both the measurements and the
calculations can be done with high precision, and the theories predict a
rich structure to the spatial power spectrum.  A multi-parameter fit of
theory to data, assuming that the fit is good, then allows simultaneous
constraints on the model parameters.
It is important to understand however that the predictions for cosmological
parameter estimation depend both on the assumed theory and on the parameter
space which one searches.  As an example, for the {\it MAP\/} satellite
scheduled to launch next year, a fit to a 7 parameter family $\Lambda$CDM
model gives errors on $\Omega_{\rm B}h^2$ of 4\%, $\Omega_{\rm mat}h^2$
of 7\%, $\Omega_\Lambda$ of 14\% and the optical depth to reionization of
14\%.
The tensor-to-scalar ratio is essentially unconstrained, as is the tensor
spectral index.
In combination these constraints, and the assumed spatial flatness, allow a
constraint on the Hubble constant of 14\%.  If we allow {\it both\/} curvature
and a cosmological constant then the error on $\Omega_\Lambda$ goes up by a
factor of 2 and on the Hubble constant by 4!

Much of the effort in parameter estimation of late has focussed on
numerical issues (where much earlier work was deficient \cite{EHT}),
on combining CMB observations with other measurements and on extending
the parameter space \cite{GDM}.
In addition to highlighting the promise of near future CMB missions, the
work on parameter estimation elucidates the often complex interplay of
cosmological parameters on the detailed structure of the anisotropy spectra.
In this regard the work on extending the parameter space is very important,
since it allows one to explore in detail the relationships that exist in
our ``favored'' models that may not exist in general.
If we can find parameters which move us off our surface of preferred theories
in a controlled manner, at the very least we can constrain such departures
when the data become available, strenghtening our belief in the fundamental
paradigm \cite{HuEis}.

The other area of much recent interest is the combination of CMB data with
the many other areas of astrophysics experiencing rapid growth.
As an example of the power of the CMB, combined with other measurements,
and of extending the parameter space, let us consider a possible measurement
of the fluctuations in the $2K$ neutrino background (here the 3 neutrino
species are assumed massless -- see Ref.~\cite{HET} for neutrino mass effects).
The hot big bang model predicts that this background must be present, and it
too will have a fluctuation spectrum.
Detecting this neutrino signal directly is almost impossible, and detecting
the {\it fluctuations\/} in the neutrino signal even more so.
However the fluctuations should be there, and their form can be predicted.

\begin{figure}
\begin{center}
\leavevmode
\epsfxsize=6cm \epsfbox{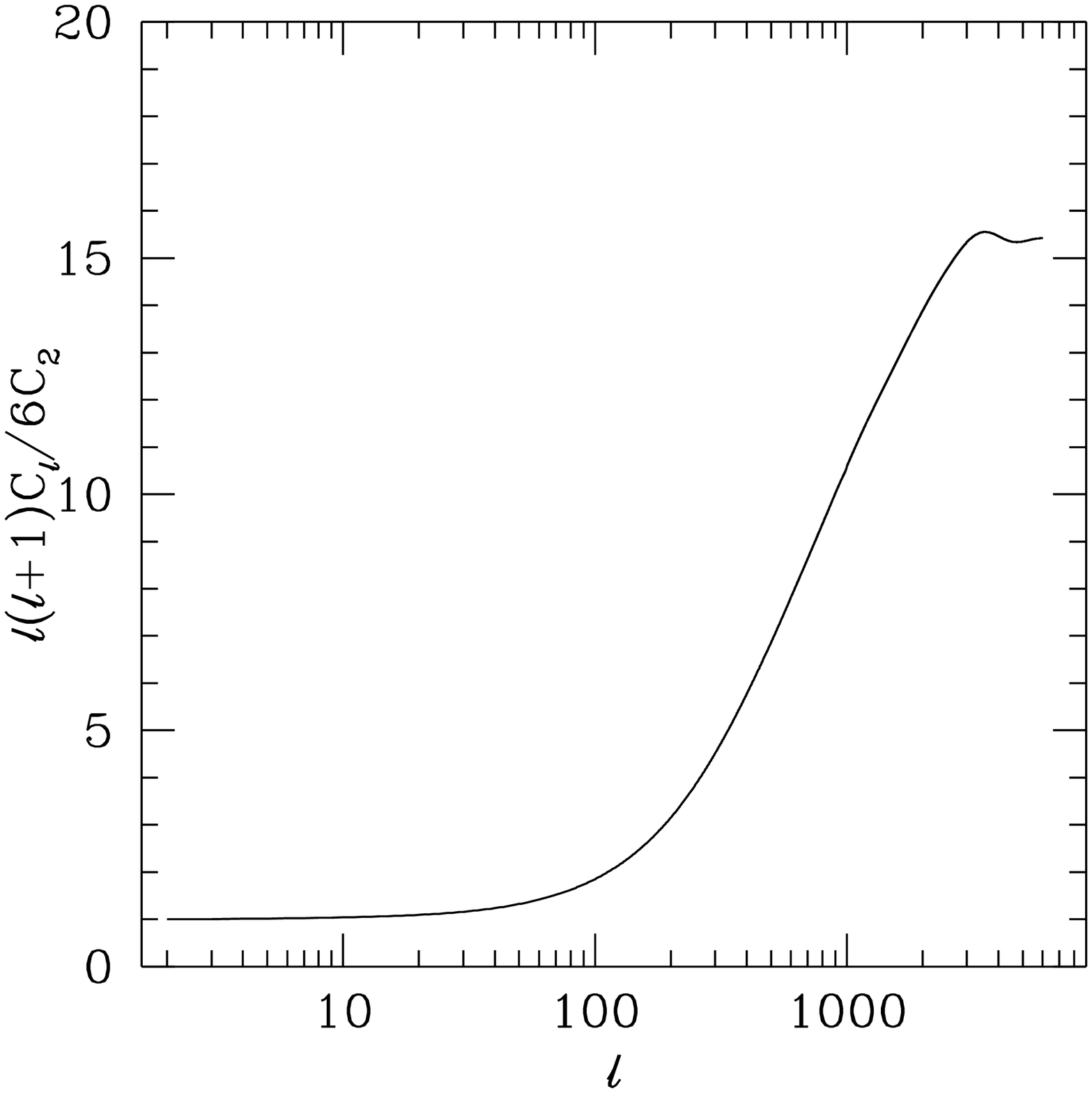}
\epsfxsize=5.5cm \epsfbox{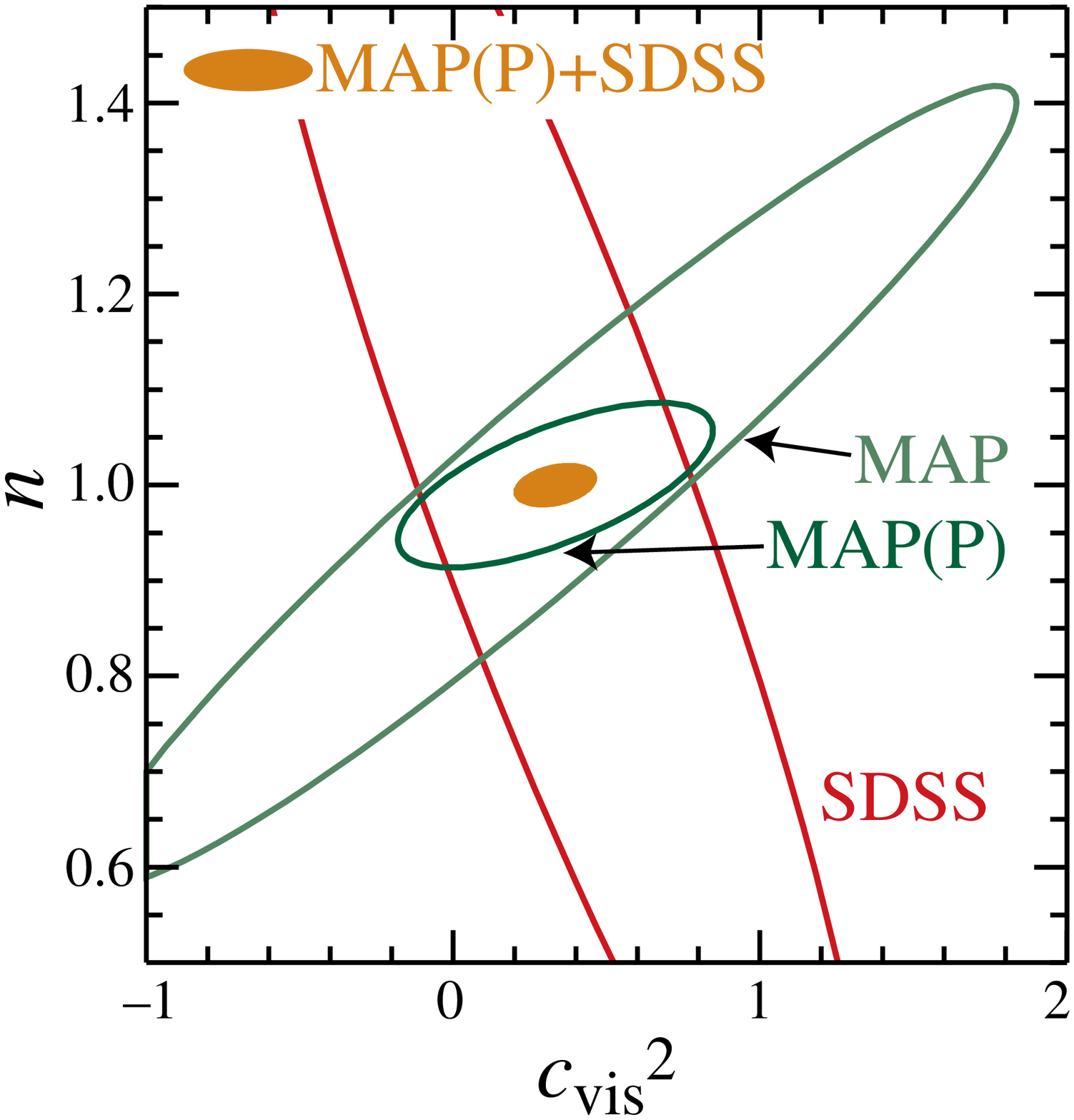}
\end{center}
\caption{(left) The anisotropy spectrum of the $2K$ cosmic neutrino background.
The small quadrupole, $\ell=2$, anisotropy in the neutrino background acts
as a source of viscosity in the fluid.  At early times when the neutrino
energy density was higher than today, this viscosity altered the evolution
of the gravitational potentials and  hence the CMB anisotropy.
(right) A possible future $1\sigma$ ``detection'' of the neutrino signal.
The ellipses show expected $1\sigma$ error contours from the experiments
listed.  The $x$-axis is the viscosity speed of sound.
The difference between $0$ and $1/3$ in $c_{\rm vis}^2$ is the signal of
the $\ell=2$ anisotropy in the neutrino spectrum of the left panel.}
\label{fig:neutrino}
\end{figure}

In the left panel of Fig.~\ref{fig:neutrino} we show a calculation of the
anisotropy spectrum, from Ref.~\cite{HSSW}.
The $y$-axis scale is $\Delta T/T\sim 10^{-5}$ as for the photons.
In the right panel of Fig.~\ref{fig:neutrino} \cite{HETW} we show a possible
future $1\sigma$ ``detection'' of the fluctuations in the neutrino background
inferred from a combination of large-scale structure and CMB data.
Clearly this particular example is somewhat fanciful.  The ``detection'' is
extremely marginal and the assumptions going into the calculation quite
optimistic.  However considering how hard it is to do this detection any
other way, it serves to illustrate the power of combinations of astrophysical
measurements to constrain fine details of all the components making up the
energy density of the universe.

\begin{figure}
\begin{center}
\leavevmode
\epsfxsize=12cm \epsfbox{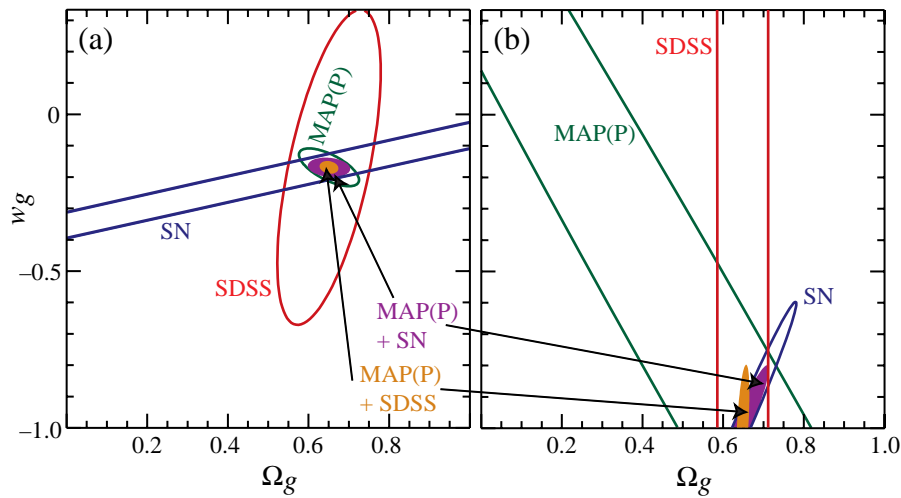}
\end{center}
\caption{The expected $1\sigma$ contours on the equation of state
$w=p/\rho$ and fraction of critical density $\Omega_g$ of a generalized
dark energy candidate.  Note that the errors do not depend strongly on
the assumed equation of state.  A model with $w_g=-1$ (right panel) is
indistinguishable from a cosmological constant.}
\label{fig:xcdm}
\end{figure}

As another (topical) example, I show in Fig.~\ref{fig:xcdm} (also taken
from Ref.~\cite{HETW}) the $1\sigma$ limits on the equation of state and
fraction of critical density in ``dark energy'' such as a cosmological
constant or dynamical scalar field (sometimes known as $x$CDM or quintessence).
Note that regardless of the equation of state a $<10\%$ measurement of both
the energy density and equation of state is possible.

These somewhat random examples should illustrate the power of the CMB to
constrain cosmological parameters, under the assumption that our current
models provide a good fit to the data.
Of course it may always turn out that while the paradigm within which we
are working is correct, our models are deficient in some detail which
prevents a good fit to the data.
The strategy in this case is to relax our assumptions and try to reconstruct
the model from the observed spectrum.  Perhaps eventually the missing
ingredient can be found and utopia regained.
In the meantime all is not lost.

\begin{figure}
\begin{center}
\leavevmode
\epsfxsize=7cm \epsfbox{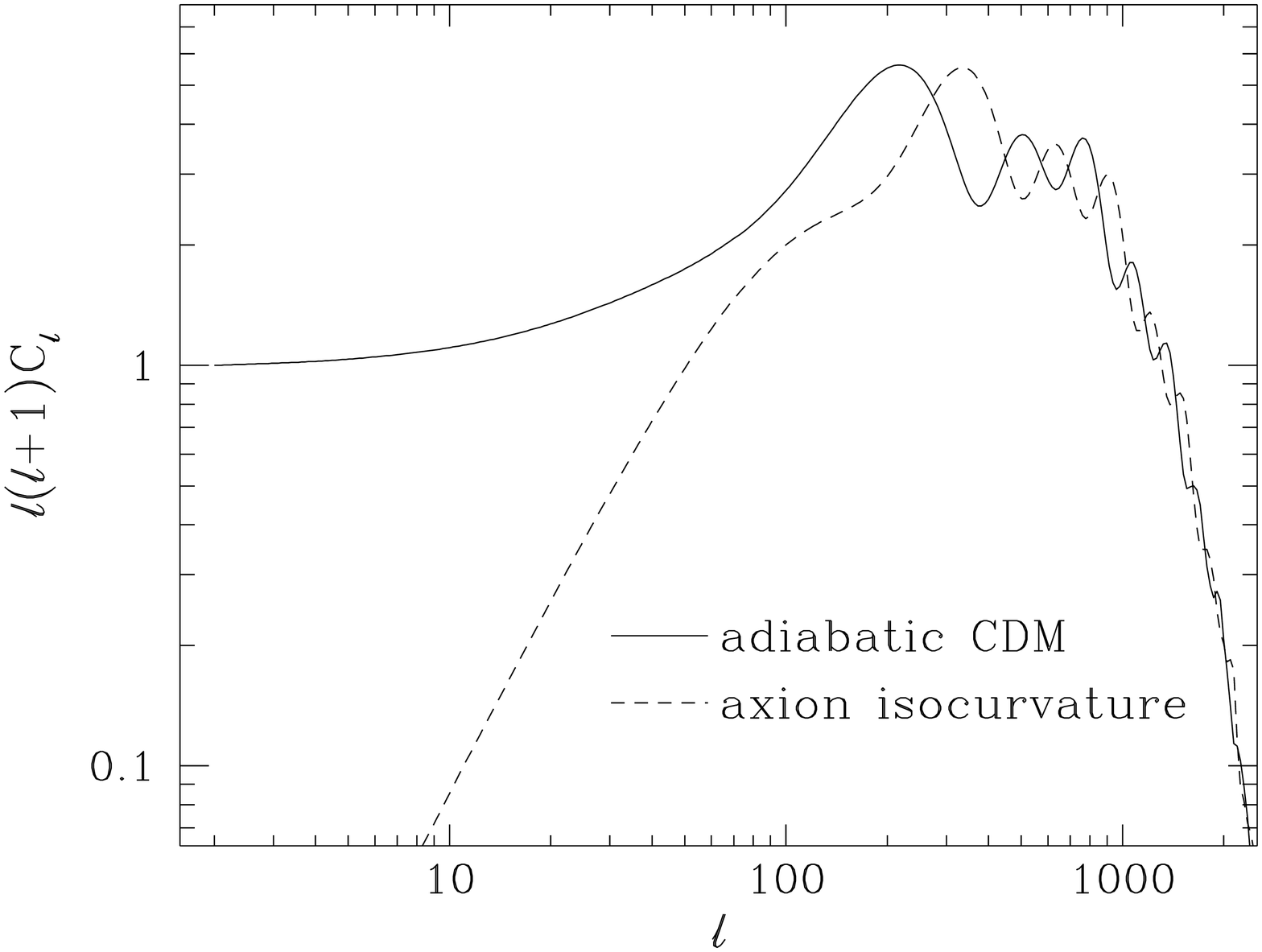}
\epsfxsize=7.2cm \epsfbox{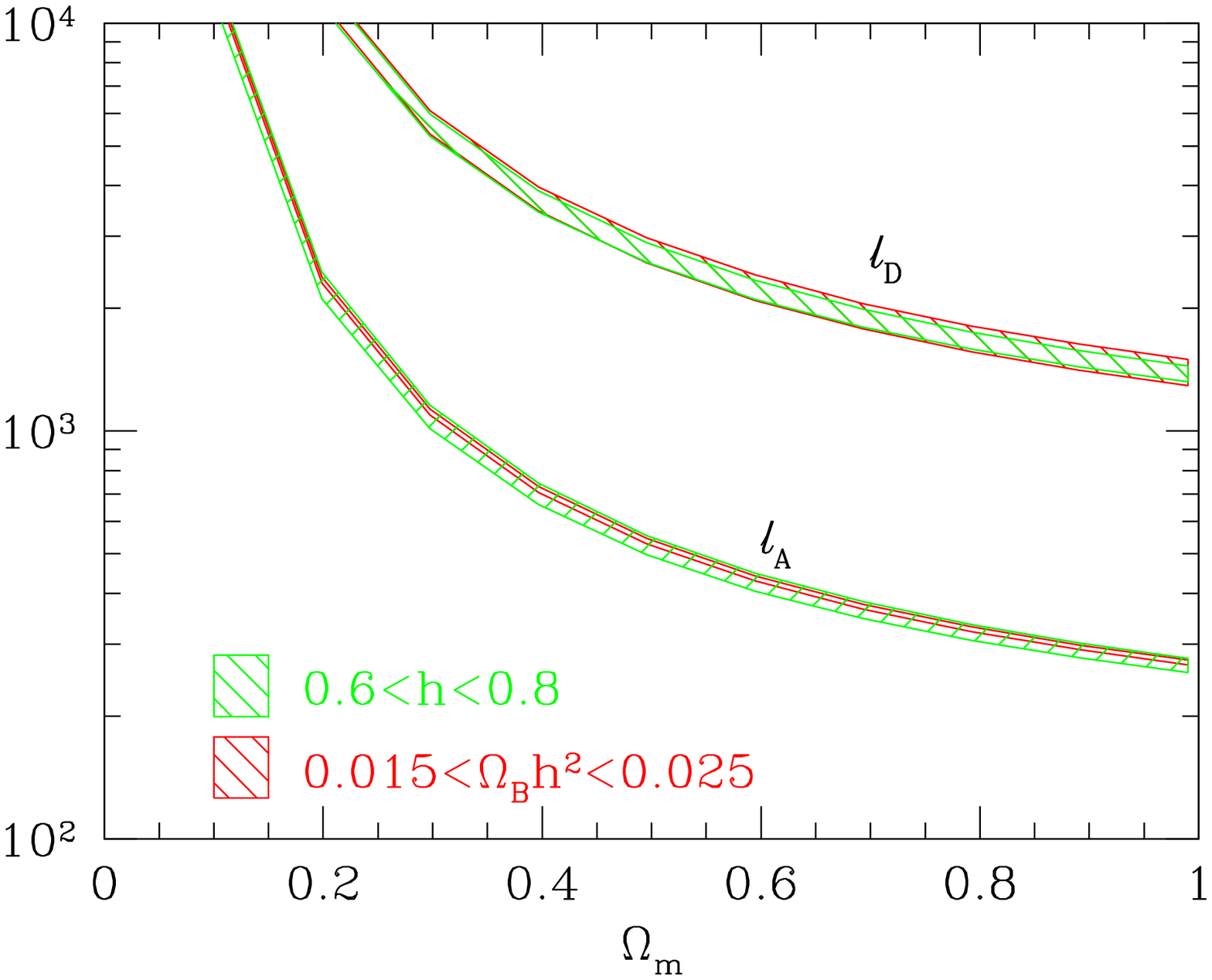}
\end{center}
\caption{(left) The anisotropy spectrum for two different models of
structure formation: an adiabatic model (solid line) and an isocurvature
model (dashed line).  Notice that the damping scale and the distance
between the 2nd and 3rd peaks is model independent.
(right) the dependence of the damping scale $\ell_D$ and the acoustic
peak scale $\ell_A$ on the angular diameter distance back to $z\sim 10^3$.
In this example, as the universe is made more open by lowering $\Omega_m$
without introducing a cosmological constant, the structure shifts to smaller
scales or higher $\ell$.}
\label{fig:modelind}
\end{figure}

There exist several model-independent measurements of the cosmological
parameters \cite{Signatures}.  As an example I show in Fig.~\ref{fig:modelind}
two different models of structure formation, both in a critical density
universe.  The models are chosen not to be good fits to the data, but to
be very different from each other.  In relativistic perturbation theory there
are two kind of perturbations: adiabatic and isocurvature.  Any fluctuation
can be decomposed into these two basis modes.  The solid line in the left
panel of Fig.~\ref{fig:modelind} is an example of a pure adiabatic model while
the dashed line is an example of a pure isocurvature model.
Note that while many things are different in these two models, there are
two things that remain fixed.  First the damping tail is at the same angular
scale in both models ($\ell\sim 1500$).  Secondly the separation between
e.g.~the 2nd and 3rd peaks is the same in both models.
The first statement is easy to understand.  The damping comes from photon
diffusion during the time it takes the universe to recombine
\cite{Silk,EfsBon87,Damping}.
Perturbations on scales smaller than the photon diffusion scale are erased,
leading to the damping of power at high-$\ell$.
Clearly this process is independent of the source of the fluctuations.
The second statement is also easy to understand.  The photon-baryon fluid
behaves like an oscillator with a natural frequency.  Once the ``bell'' is
struck it wants to ring at that natural frequency.  Thus even though the
driving forces in the adiabatic and isocurvature models are different, the
``ringing'' of the higher peaks proceeds at the same frequency.  So the
peak spacing is fixed.

While the peak spacing $\ell_A$ and the damping scale $\ell_D$ are (nearly)
independent of the model of structure formation, they do depend on the
cosmology.  Specifically on the mapping between physical scales at the
surface of last scattering ($z\sim 10^3$) and angles on the sky.  They are
thus probes of the angular diameter distance to last scattering, as shown
in the right panel of Fig.~\ref{fig:modelind} for the case of an open
universe.  More general constraints in the $\Omega_m-\Omega_\Lambda$ plane
or the $\Omega_g-w$ plane can be found in \cite{Comp,HETW} respectively.

To turn the problem around one can look for tests of the model of structure
formation independent of the cosmology.
Our most succesful class of models is those with an early epoch of accelerated
expansion, i.e.~inflationary models.  Since accelerated expansion requires a
fluid with negative pressure, it is intimately related to quantum mechanical
considerations (the inner space--outer space connection).
One of the greatest triumphs of the inflationary idea is that it provides a
source of small adiabatic fluctuations which can grow, through gravitational
instability, to form the CMB anisotropies and large-scale structure that we
observe today.
How can we test this paradigm for the generation of primoridal fluctuations?
Any model of fluctuations should produce all 3 modes of perturbations:
scalar modes (density perturbations), vector modes (fluid vorticity) and
tensor modes (gravitational waves).  The vectors have no growing mode and
so after a few expansion times they have decayed away, leaving scalar and
tensor modes.  The presence of vector modes would thus be evidence for
fluctuation generation activity while the CMB anisotropy was being formed,
i.e.~not inflation.  The mere presence of tensor modes does not however argue
one way or the other.

In inflationary models based on a single, slow-rolling scalar field the scalar
modes are enhanced over the tensor modes by a large factor which is related
to the {\it tensor\/} spectral index (see Ref.~\cite{InfRev} for a review).
The relation becomes an inequality if more than one field is important.
Unfortunately the tensor spectral index is quite hard to measure unless the
tensor signal is large, and usually an additional (model dependent) relation
to the {\it scalar\/} spectral index is assumed instead.
It has been argued that if the inflationary idea is to find a home in modern
high-energy physics theories, rather than in effective or ``toy'' models, then
the tensor signal is quite likely to be small \cite{Lyth}.
While our ignorance of physics above the electroweak scale makes it dangerous
to take particle physics predictions as gospel in cosmology, the observational
situation also argues against a large tensor signal \cite{ZibScoWhi}.
In some sense this is good news: a generic mechanism would presumably make
scalar, vector and tensor perturbations in roughly the same amounts leading
to $T/S\simeq 1$ today (the vectors having decayed).  Inflation on the other
hand predicts that the scalar signal is enhanced, lowering $T/S$ from this
naive prediction, as observations currently prefer.

Luckily CMB based tests of inflation exist which do {\it not\/} require a
measurement of the tensor signal \cite{TestInf,HSW}.
They rely on the fact that the only known way to generate {\it adiabatic\/}
fluctuations (i.e.~fluctuations in the energy density or curvature of space)
on cosmological scales today is to have a period of accelerated expansion
\cite{HuTurWei,Liddle,Signatures}, i.e.~inflation.
The key then is to test for the adiabaticity of the fluctuations, which can
be done with broader features than detailed fitting to extract small signals.
Plausibility arguments suggest that if a peak in the anisotropy spectrum is
observed near $\ell\sim 200$, the fluctuations are adiabatic.
Isocurvature models generically predict a peak shifted to higher $\ell$
\cite{Signatures,HSW} (see Fig.~\ref{fig:modelind}).
Further support for this inference could be gained by measuring the 2nd and
3rd peaks, though some loopholes still remain
\cite{Signatures,Turok1,Turok2,HSW}.
The sharpest tests of the model \cite{HSW} can be performed if information
about the polarization of the CMB is obtained (as both MAP and Planck intend).

\begin{figure}
\begin{center}
\leavevmode
\epsfxsize=6cm \epsfbox{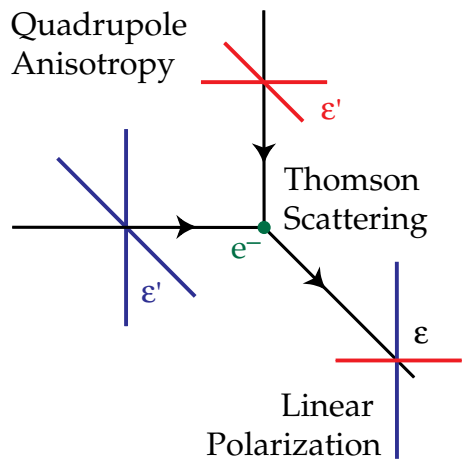}
\end{center}
\caption{A quadrupole anisotropy at the surface of last scattering
generates a linear polarization, since the angular dependence of Thomson
scattering depends on the polarization of the ingoing and outgoing radiation.}
\label{fig:poln}
\end{figure}

Since $\gamma-e$ scattering depends on polarization and angle as
$\epsilon_f\cdot\epsilon_i$, where $\epsilon_{f,i}$ are the polarization
vectors of the final and inital radiation, a quadrupole anisotropy
generates linear polarization (see Fig.~\ref{fig:poln}).
An introduction to polarization can be found in Ref.~\cite{Polar}, and the
numerous references therein.
For our purposes here the key feature of polarization is that it is
generated {\it only\/} by scattering.  The small angle polarization is thus
localized to the last-scattering surface, and provides us with a probe of
the anisotropies (as a function of scale) at that time.
The behaviour of the anisotropy around the horizon scale, and the slope of
the spectrum at larger scales, then gives a test of the presence or absence
of large-scale fluctuations in the curvature \cite{TAMM,Polar}.

In conclusion, cosmology is now in a ``golden age''.  We finally have the
data to answer our most fundamental questions, and to generate new puzzles.
Within a decade we hope to have a standard model of structure formation.
Our current theoretical structure, starting with quantum fluctuations in the
early universe, continuing with general relativistic dynamics and ending
with free-fall of radiation and matter, is one of the most beautiful and
complete in all of physics.  Far from the cosmology of old, where order of
magnitude estimates held sway, modern cosmology emphasizes precision
calculations using well controlled approximations.
The archetypical system of this ``new era'' is the microwave background.
If our models are close to correct, high precision studies of the CMB
anisotropy will revolutionize cosmology.  If our models are wrong, one
could not hope for a better data set with which to find the right path.
We are all eagerly awaiting imminent experimental advances in this field.

\end{document}